\pdfoutput=1
\documentclass[conference]{IEEEtran}
\IEEEoverridecommandlockouts
\usepackage{cite}
\usepackage{amsmath,amssymb,amsfonts}
\usepackage{algorithm}
\usepackage{algorithmic}
\usepackage{graphicx}
\usepackage{textcomp}
\usepackage{xcolor}
\ifCLASSINFOpdf
\else
\fi
\begin{document}
%
\title{DMA-Aided MU-MISO Systems for Power Splitting SWIPT via Lorentzian-Constrained Holography}

\author{Askin Altinoklu, and Leila Musavian
\thanks{The authors are with the School of Computer Science and Electronic Engineering, University of Essex, Wivenhoe Park, Colchester CO43SQ, United Kingdom (e-mail: {askin.altinoklu, leila.musavian}@essex.ac.uk). This work was supported by UK Research and Innovation under the UK government’s Horizon Europe funding guarantee through MSCA-DN SCION Project Grant Agreement No.101072375 [Grant Number: EP/X027201/1].}}

%


\maketitle

\begin{abstract}
This paper presents an optimal power splitting and beamforming design for co-located simultaneous wireless information and power transfer (SWIPT) users in Dynamic Metasurface Antenna (DMA)-aided multiuser multiple-input single-output (MISO) systems. The objective is to minimize transmit power while meeting users’ signal-to-interference-plus-noise ratio (SINR) and energy harvesting (EH) requirements. The problem is solved via an alternating optimization framework based on semi-definite programming (SDP), where metasurface tunability follows Lorentzian-constrained holography (LCH). In contrast to traditional beamforming architectures, DMA-assisted architectures reduce the need for RF chains and phase shifters; however, they require optimization under Lorentzian constraints that couple amplitude and phases. Hence, the proposed method integrates several LCH schemes, including the recently proposed adaptive-radius LCH (ARLCH), and evaluates nonlinear EH models and circuit noise effects. Simulation results show that the proposed design significantly reduces transmit power compared with baseline methods, highlighting the efficiency of ARLCH and optimal power splitting in DMA-assisted SWIPT systems.
\end{abstract}


%
\IEEEpeerreviewmaketitle

\section{Introduction}
DMAs, a recently emerging class of holographic multiple-input multiple-output (HMIMO) systems, consist of reconfigurable arrays of microstrip- or waveguide-fed metasurface elements whose electromagnetic responses can be dynamically controlled by tuning parameters for changing the resonant state of elements \cite{em_dma2}. This tunability enables analog-domain beamforming without the need for external phase shifters or high-power amplifiers that are typically required in traditional hybrid or fully digital architectures \cite{DMA_BF1}. When integrated with digital precoders, DMAs can establish multiuser MIMO links with fewer RF chains, thereby enabling scalable and energy-efficient transceiver architectures \cite{DMA1}. 

SWIPT is another key technology for next-generation communication systems, allowing the simultaneous delivery of wireless information transfer (WIT) and wireless power transfer (WPT) \cite{survey_WIET_6G}. Prior research on SWIPT in traditional MIMO architectures has focused on improving end-to-end energy transfer and spectral efficiency in co-located receivers via power-splitting (PS) and time-switching (TS) schemes \cite{MIMO_PS_SWIPT}. On the other hand, for networks with separated energy-harvesting (EH) and information-decoding (ID) users, beamforming-based approaches were employed to allocate dedicated transmission beams \cite{MIMO_Sep_SWIPT}. However, the increasing demand for energy and information transfer in future generation systems calls for  energy-efficient transceiver architectures, where DMAs can play a significant role due to their hardware simplicity and beamforming flexibility. Despite growing interest, there are only a few works considering DMA-assisted SWIPT systems. Existing works have mainly focused on either WPT \cite{DMA_Nearfield_WPT} or WIT based \cite{DMA_BF1,aaltinoklu} designs, particularly in near-field communications. Recent studies, such as \cite{DMA_app6} and \cite{DMA_SWIPT2} considered DMA-aided SWIPT for spectral efficiency and harvested power maximization, respectively. however, both addressed separate EH and ID users, achieved through separate beamforming vectors. In contrast, the co-located SWIPT scenario, where each user simultaneously performs EH and ID remains largely unexplored for DMA-aided architectures, despite being highly relevant to low-power wireless-powered sensor and internet of things networks in 6G use-cases.

The design of DMA-assisted PS-SWIPT systems is particularly challenging due to the need for joint optimization of DMA weights, digital precoders, and PS factors. Moreover, DMA weights are governed by Lorentzian constraints, where the amplitude of each metasurface element is intrinsically coupled to its phase \cite{em_dma2}. In particular, achieving holographic beamforming with DMAs requires mapping ideal beamforming weights of the form $e^{j\phi}$ to physically realizable Lorentzian-constrained responses, which inherently follow the structure $\frac{j+e^{j\phi}}{2}$. To address this constraint, several LCH schemes have been proposed, including phase holography (LCPH), Euclidean holography (LCEH), unitary-shift holography (LCUSH), and amplitude-only holography (AOH) \cite{DMA3, aaltinokluLCH}. Existing optimization approaches for DMA beamforming include manifold-based methods, alternating direction method of multipliers (ADMM), and majorization--minimization techniques under various WIT and SWIPT objectives \cite{DMA_BF1, Yihan_Near, DMA_app6}. Although these techniques offer lower computational complexity, they are typically tailored to a specific LCH realization, most commonly LCUSH, and do not readily generalize to other LCH schemes. To allow systematic evaluation and fair comparison between multiple LCH strategies, a generalized optimization framework based on SDP was recently introduced in \cite{aaltinokluLCH}, integrating LCPH, LCEH, and LCUSH within a unified convex formulation. Building on this framework, the ARLCH scheme was proposed to dynamically optimize the Lorentzian projection and improve the amplitude--phase tradeoff. However, the application of such a unified SDP-based framework and the ARLCH scheme to SWIPT systems has not yet been investigated. This extension is particularly challenging due to the coexistence of SINR and EH constraints, the introduction of additional optimization variables associated with PS factors, and the impact of practical nonlinear energy-harvesting models.

The main contributions of this paper are summarized as follows:
\begin{itemize}
    \item We propose a unified optimization framework for DMA-assisted PS-SWIPT in multiuser MISO downlink systems, formulating a transmit power minimization problem under simultaneous SINR and EH constraints. The resulting nonconvex problem is relaxed into an SDP form, which serves as a tractable and unifying benchmark for DMA beamforming. 
    \item To the best of our knowledge, this is the first work to jointly address DMA weight realization, digital precoding, and power-splitting design for co-located SWIPT. By leveraging the unified SDP-based benchmark, we systematically evaluate multiple LCH mapping strategies and integrate the ARLCH scheme into the DMA-SWIPT setting, enabling improved amplitude-phase tradeoff and reduced Lorentzian projection loss.
    \item Unlike existing DMA-SWIPT studies, the proposed formulation includes practical nonlinear energy-harvesting models and we examine their impact on beamforming performance and transmit power efficiency.
    \item Numerical results demonstrate that the proposed ARLCH-based PS-SWIPT approach achieves significant transmit power reductions compared to benchmark LCH schemes while satisfying both SINR and EH requirements. The results further reveal that nonlinear EH models and optimal power splitting have a pronounced impact on the required transmit power.
\end{itemize}

\textbf{Notations:} Matrices are denoted by \(\mathbf{W}\), with  \(\text{Tr}(\mathbf{W})\), \(\mathbf{W}^T\), \(\mathbf{W}^H\), and \(\text{Vec}(\mathbf{W})\) representing, trace, transpose, Hermitian conjugate, and vectorization. Vectors are \(\mathbf{w}\), with \(\|\mathbf{w}\|\), \(\mathbf{w}^T\), and \(\mathbf{w}^H\) indicating Euclidean norm, transpose, and Hermitian conjugate. Scalars are \(w\), with \(|w|\) as the absolute value. The Kronecker product is denoted by \(\otimes\).

\section{System Model and Problem Formulation}
We consider a multiuser MISO downlink SWIPT system comprising a base station (BS) equipped with a DMA-based hybrid architecture integrated with a digital precoder. The network serves multiple user equipments (UEs), each of which requires a target SINR for information decoding and a minimum harvested energy level.

\subsection{Channel and Path-Loss Modeling}
In this paper, we consider a spherical wavefront model (SWM) under near-field line-of-sight (LoS) propagation.
For obtaining SWM, we consider a uniform planar array (UPA) consisting of \(N_\text{r}\) and \(N_\text{c}\) antenna elements in the vertical and horizontal directions, yielding a total of \(N = N_\text{r}N_\text{c}\) elements. 
For this model, the radiation pattern of the elements is well-approximated with:
\begin{equation}
\label{eq:1}
G_\text{e}(\psi) = 
\begin{cases} 
6 \cos^{2}(\psi), & 0 \leq \psi \leq \pi/2, \\
0, & \pi/2 < \psi \leq \pi,
\end{cases}
\end{equation}
where $\psi$ is the angle measured from the array bore-sight\cite{gain_element}. 
We consider a multiuser downlink system with $K$ users, indexed by $k=1,\ldots,K$.
The path loss for the $\mathrm{UE}_k$ positioned at $\mathbf{r}_k$ due to the ($i$,$l$)-th element of the UPA (\(\mathbf{r}_{i,l}\)) can be expressed with:
\begin{equation}
\label{eq:2}
\gamma_{k}(i, l) = \sqrt{G_\text{e}(\psi)} \frac{\lambda}{4\pi \|\mathbf{r}_k - \mathbf{r}_{i, l}\|} e^{-j \beta_0 \|\mathbf{r}_k - \mathbf{r}_{i, l}\|},
\end{equation}
where $\lambda$ and $\beta_0$ denote the free-space wavelength and wavenumber, respectively. The indices $i=1,\ldots,N_{\mathrm{r}}$ and $\ell=1,\ldots,N_{\mathrm{c}}$ correspond to the vertical and horizontal array elements, respectively. The entries \(\gamma_{k}(i, l)\) are the elements of the channel vector \(\boldsymbol{\gamma}_k \in \mathbb{C}^{N \times 1}\), i.e., \(\boldsymbol{\gamma}_k \triangleq \begin{bmatrix}
\gamma_{k}(1, 1), \gamma_{k}(1, 2), \ldots, \gamma_{k}(N_\text{r}, N_\text{c})
\end{bmatrix}^H\).
\subsection{DMA Architecture}
The reference waves traveling within each DMA radiate through the metasurfaces whose weights are externally controlled by dynamically configurable weights based on LCH (\(\mathbf{Q} \in \mathbb{C}^{N \times N_\text{r}}\)) to achieve analog beamforming. Each DMA is connected to the digital precoder via RF chains, and the weights of reference waves of different DMAs are further controlled by the information encoded linear precoding vectors \(\mathbf{w}_m \in \mathbb{C}^{N_\text{r} \times 1}\) for all \(m \in M\), where \(M = \min(K, N_\text{r})\). The propagation of reference waves within the DMAs are captured by \(\mathbf{H} \in \mathbb{C}^{N \times N}\), which is a diagonal matrix with elements \(\mathbf{H}_{(i-1)N_\text{c} + l, (i-1)N_\text{c} + l} = e^{-d_{i,l} (\alpha_i + j \beta_i)}\), where \(d_{i,l}\) is the position of the \(l\)-th element along the \(i\)-th microstrip, and \(\alpha_i, \beta_i\) are its attenuation and propagation constants \cite{DMA1}. Based on this, the baseband transmitted signal at DMA-aided BS is expressed as
\begin{equation}
\mathbf{x} = \sum_{m=1}^{M} \mathbf{x}_m = \sum_{m=1}^{M} \mathbf{H} \mathbf{Q} \mathbf{w}_m s_m,
\label{eq:3}
\end{equation}
where \(s_m\) denotes the transmitted data symbol associated with the \(m\)-th precoding vector \(\mathbf{w}_m\), for all \(m \in \{1,\ldots,M\}\).
In \eqref{eq:3}, the matrix \(\mathbf{Q}\) has a block-diagonal structure, where each block represents the amplitude–phase profile of the metasurface along a distinct DMA microstrip, i.e., each row of the UPA, and is given by
\begin{equation}
\label{eq:4}
\mathbf{Q}_{(i-1)N_\text{c} + l, n} =
\begin{cases}
q_{i,l} \in \mathbb{Q},  & i = n,\\
0, & i \neq n,
\end{cases}
\end{equation}
where \(\mathbb{Q}\) denotes the Lorentzian-profile set characterizing the tunable metasurface elements.  
Each coefficient \(q_{i,l}\) is controlled by varying the resonance frequency of its associated element, for example via external varactor diodes, and satisfies
\begin{equation}
\label{eq:5}
\mathbb{Q} = \left\{ q = \frac{j + e^{j\Phi}}{2} : \Phi \in [0, 2\pi] \right\}.
\end{equation}
\subsection{SWIPT Signal Modeling}
In the considered PS-SWIPT scenario, the received RF signal at \(\mathrm{UE}_k\) is split by a power splitter into two parts: a fraction \(\rho_k\) (\(0<\rho_k<1\)) directed to the information decoder (ID) and the remaining \(1-\rho_k\) directed to the energy harvester (EH). Accordingly, the signal at the ID is given by
\begin{equation}
\label{eq:7}
y_k^{(\mathrm{ID})}
   = \sqrt{\rho_k}\!\left(\boldsymbol{\gamma}_k^{H}\!\sum_{m=1}^{M}\mathbf{H}\mathbf{Q}\mathbf{w}_m s_m + n_{\mathrm{a},k}\right)
     + n_{\mathrm{c},k},
\end{equation}
where \(n_{\mathrm{a},k}\sim\mathcal{CN}(0,\sigma_{\mathrm{a},k}^{2})\), and \(n_{\mathrm{c},k}\sim\mathcal{CN}(0,\sigma_{\mathrm{c},k}^{2})\) represent the antenna noise at \(\mathrm{UE}_k\), and the additional noise introduced by baseband conversion in the ID of \(\mathrm{UE}_k\), respectively.  
Using \eqref{eq:7}, the SINR at \(\mathrm{UE}_k\) is given by
\begin{equation}
\label{eq:SINR}
\mathrm{SINR}_k
 = \frac{\bigl|\boldsymbol{\gamma}_k^{H}\mathbf{H}\mathbf{Q}\mathbf{w}_k\bigr|^{2}}
        {\displaystyle\sum_{\substack{m\neq k}}^{M}
          \bigl|\boldsymbol{\gamma}_k^{H}\mathbf{H}\mathbf{Q}\mathbf{w}_m\bigr|^{2}
          + \sigma_k^{2}},
\end{equation}
where \(\sigma_k^{2} = \sigma_{\mathrm{a},k}^{2} + \sigma_{\mathrm{c},k}^{2}/\rho_k\).
Similarly, for the EH part, the received signal at \(\mathrm{UE}_k\) is expressed by 
\begin{equation}
\label{eq:8}
y_k^{(\mathrm{EH})}
   = \sqrt{1-\rho_k}\!\left(\boldsymbol{\gamma}_k^{H}\!\sum_{m=1}^{M}\mathbf{H}\mathbf{Q}\mathbf{w}_m s_m + n_{\mathrm{a},k}\right),
\end{equation}
while the received power for the EH part is obtained with 
\begin{equation}
\label{eq:9}
P_{k}^{(\mathrm{EH})}
   = (1 - \rho_k)\sum_{m=1}^{M}
     \left| \boldsymbol{\gamma}_k^{H}\mathbf{H}\mathbf{Q}\mathbf{w}_m \right|^{2}.
\end{equation}
Then, under the linear EH model, the harvested energy at \(\mathrm{UE}_k\) is linearly proportional to the received RF power, i.e., \(E_k = \eta P_k^{(\mathrm{EH})}\), where \(0 < \eta \le 1\) denotes the energy-harvesting efficiency.
For a better representation, the harvested energy can alternatively be described by a logistic nonlinear model \cite{nonlinear_harvester}:
\begin{equation}
\label{eq:nonlinear_EH_models}
E_k =
\dfrac{\dfrac{E^{\text{sat}}}{1+e^{-a\,(P_k^{(\mathrm{EH})}-b)}}
      - \dfrac{E^{\text{sat}}}{1+e^{a b}}}
      {1 - \dfrac{1}{1+e^{a b}}},
\end{equation}
where \(E^{\text{sat}}\) is the maximum harvestable power before saturation and \(a\) and \(b\) are circuit-specific logistic parameters.

\section{Problem Formulation and Solution}
\subsection{Optimization Problem Formulation}
In this paper, we consider a PS-SWIPT system under guaranteed QoS requirements for both ID and EH.  
The objective is to minimize the total transmit power at the DMA-aided BS by jointly designing the precoding vectors \(\mathbf{w}_k\), the DMA weight matrix \(\mathbf{Q}\), and the PS ratios \(\rho_k\).  
The resulting optimization problem is formulated as
\begin{subequations}
\label{eq:10}
\begin{flalign}
\underset{\{\mathbf{w}_m\}_{k=1}^K,\{\rho_k\}_{k=1}^K,\mathbf{Q}}{\text{minimize}}
&\quad \sum_{k=1}^{K}\bigl\|\mathbf{H}\mathbf{Q}\mathbf{w}_k\bigr\|^{2} && 
\label{eq:10a}\\[2pt]
\text{s. t.}\quad
& \mathrm{SINR}_k(\{\mathbf{w}_k\},\mathbf{Q},\rho_k) \ge \delta_k,
&& \forall k,
\label{eq:10b}\\[4pt]
& E_k(\{\mathbf{w}_k\},\mathbf{Q},\rho_k) \ge E_k^{\mathrm{th}},
&& \forall k,
\label{eq:10c}\\[2pt]
& 0<\rho_k<1,
&& \forall k,
\label{eq:10d}\\[2pt]
& \mathbf{Q}\in\mathbb{Q}^{N\times N_{\mathrm{r}}}, &&
\label{eq:10e}
\end{flalign}
\end{subequations}
where $\delta_k$, and $E_k^{\mathrm{th}}$ are the SINR and EH thresholds of $\mathrm{UE}_k$, respectively. Constraint \eqref{eq:10c} can be equivalently expressed using the received RF power from \eqref{eq:9} for both the linear and nonlinear energy-harvesting models as
\begin{equation}
\label{eq:EH_models}
P_{k}^{(\mathrm{EH})} \ge P_k^{\mathrm{th}},
\end{equation}
where
\begin{equation*}
P_k^{\mathrm{th}} =
\begin{cases}
\dfrac{E_k^{\mathrm{th}}}{\eta}, &
  \text{(Linear model)},\\[6pt]
b - \dfrac{1}{a}\ln
    \dfrac{e^{a b}\bigl(E^{\text{sat}}-E_k^{\mathrm{th}}\bigr)}
          {e^{a b}E_k^{\mathrm{th}} + E^{\text{sat}}}, &
  \text{(Nonlinear model)}.
\end{cases}
\end{equation*}

Moreover, the constraint \eqref{eq:10b} can be written as:
\begin{equation}
\bigl|\boldsymbol{\gamma}_k^{H}\mathbf{H}\mathbf{Q}\mathbf{w}_k\bigr|^{2}
 - \delta_k\sum_{\substack{j=1\\ j\ne k}}^{K}\bigl|\boldsymbol{\gamma}_k^{H}\mathbf{H}\mathbf{Q}\mathbf{w}_j\bigr|^{2}
 \ge \delta_k\bigl(\sigma_{\mathrm{a},k}^{2} + \frac{\sigma_{\mathrm{c},k}^{2}}{\rho_k}\bigr).
\label{eq:11}
\end{equation}

The optimization problem in \eqref{eq:10} involves coupled variables, namely the precoding vectors \(\mathbf{w}_k\), the DMA weight matrix \(\mathbf{Q}\), and the PS ratios \(\rho_k\).  
It involves quadratic terms that appear in a non-convex form, particularly in the coupling of \(\mathbf{Q}\) and \(\mathbf{w}_k\).  
Furthermore, the Lorentzian constraint in \eqref{eq:10e} is also non-convex, since the amplitude and phase of each DMA element are intrinsically linked through the sinusoidal relationship described in \eqref{eq:5}.  
As a result, solving this problem is challenging; in the following sections, we present a tractable solution approach. 
\subsection{Optimizing the Digital Precoder and PS Ratio}
When the DMA weight matrix \(\mathbf{Q}\) are fixed, the problem in \eqref{eq:10} can be reformulated as the following SDP:
\begin{equation}
\label{eq:12}
\begin{aligned}
\underset{\{\mathbf{W}_k\,\rho_k\}}{\text{min}}\quad
    & \sum_{k=1}^{K}\operatorname{Tr}(\mathbf{Z}\mathbf{W}_k) \\[1pt]
\text{s. t.}\quad
     \frac{1}{\delta_k}\operatorname{Tr}(\mathbf{P}_k\mathbf{W}_k)
      -  \!\!&\sum_{\substack{m=1\\ m\ne k}}^{K}
        \operatorname{Tr}(\mathbf{P}_k\mathbf{W}_m)
       \ge \sigma_{\mathrm{a}}^{2} + \frac{\sigma_{\mathrm{c}}^{2}}{\rho_k},  &&\forall k, \\[1pt]
    & \sum_{m=1}^{K}\operatorname{Tr}(\mathbf{P}_k\mathbf{W}_m)
      \ge \frac{P_k^{\mathrm{th}}}{1 - \rho_k},\quad &&\forall k, \\[1pt]
    & 0<\rho_k<1, \quad &&\forall k, \\[1pt]
    & \operatorname{rank}(\mathbf{W}_k) = 1,\quad &&\forall k, \\[1pt]
    & \mathbf{W}_k \succeq 0,\quad &&\forall k,
\end{aligned}
\end{equation}
where $\mathbf{Z}=(\mathbf{H}\mathbf{Q})^{H}\mathbf{H}\mathbf{Q}$,  
$\mathbf{P}_k=(\boldsymbol{\gamma}_k^{H}\mathbf{H}\mathbf{Q})^{H}\boldsymbol{\gamma}_k^{H}\mathbf{H}\mathbf{Q}$,  
and $\mathbf{W}_k=\mathbf{w}_k\mathbf{w}_k^{H}$. By dropping the rank-one constraint $\operatorname{rank}(\mathbf{W}_k)=1$, problem~\eqref{eq:12} is relaxed into a convex SDP. The objective is linear in $\mathbf{W}_k$, and the remaining constraints are convex in $\{\mathbf{W}_k,\rho_k\}$, as they consist of affine functions of $\mathbf{W}_k$ and convex functions of $\rho_k$ over $0<\rho_k<1$. The relaxed problem can be efficiently solved using standard tools such as CVX~\cite{cvx1}. Moreover, the relaxation can be shown to yield rank-one solutions, allowing direct recovery of the beamforming vectors $\mathbf{w}_k$ from the principal eigenvectors of $\mathbf{W}_k$ \cite{SDP}. Hence, the optimized digital precoder vectors and PS ratios can be obtained by solving this problem.

\subsection{Optimizing the DMA Weights}
To optimize the DMA weights, we adopt a two-tier procedure. For fixed given \(\mathbf{w}_k\) and \(\rho_k\), we first solve an SDP to obtain the ideal beamforming weights, and then apply an LCH mapping to project them onto the Lorentzian-feasible set.
When the digital precoding vectors $\{\mathbf{w}_k\}$ and the PS ratios $\{\rho_k\}$ are fixed, the Lorentzian-relaxed DMA weight optimization problem in \eqref{eq:10} can be reformulated into a tractable SDP form with respect to $\mathbf{Q}$ by applying standard vectorization and trace identities as following \cite{aaltinokluLCH}:
\begin{equation}
\label{eq:13}
\begin{aligned}
\underset{\mathbf{\tilde{Q}}}{\text{minimize}}\quad
    & \sum_{m=1}^{M}\operatorname{Tr}\bigl(\mathbf{\tilde{B}}_m \mathbf{\tilde{Q}}\bigr)
    \\[1pt]
\text{s.t.}\quad
     \operatorname{Tr}\bigl(\mathbf{\tilde{C}}_{k,k}\mathbf{\tilde{Q}}\bigr)
      - \delta_k\!\!&\sum_{\substack{m=1\\ m\neq k}}^{M}
        \operatorname{Tr}\bigl(\mathbf{\tilde{C}}_{k,m}\mathbf{\tilde{Q}}\bigr)
      - \delta_k \sigma_k^{2} \ge 0,&&\forall k,
     \\[1pt]
    & \sum_{m=1}^{M}
        \operatorname{Tr}\bigl(\mathbf{\tilde{C}}_{k,m}\mathbf{\tilde{Q}}\bigr)
        \ge \frac{P_k^{\mathrm{th}}}{1 - \rho_k}, &&\forall k,
     \\[1pt]
    & \mathbf{\tilde{Q}}\succeq 0.
\end{aligned}
\end{equation}
where
\(\mathbf{\tilde{B}}_m
   = \bigl(\mathbf{w}_m^{T}\!\otimes\!\mathbf{H}\bigr)^{H}
     \bigl(\mathbf{w}_m^{T}\!\otimes\!\mathbf{H}\bigr)\), and
\(\mathbf{\tilde{C}}_{k,m}
   = \bigl(\mathbf{w}_m^{T}\!\otimes\!(\boldsymbol{\gamma}_k^{H}\mathbf{H})\bigr)^{H}
     \bigl(\mathbf{w}_m^{T}\!\otimes\!(\boldsymbol{\gamma}_k^{H}\mathbf{H})\bigr)\). 
Here, $\tilde{\mathbf{Q}} = \tilde{\mathbf{q}}\tilde{\mathbf{q}}^{H}$, 
where $\tilde{\mathbf{q}}\in\mathbb{C}^{N\times 1}$ is obtained by removing the zero entries 
from $\mathbf{q}=\operatorname{vec}(\mathbf{Q})\in\mathbb{C}^{L\times 1}$, 
with $L=N_{\mathrm{r}}^{2}N_{\mathrm{c}}$. 
Accordingly, the matrices $\tilde{\mathbf{B}}_m$ and $\tilde{\mathbf{C}}_{k,m}$ 
are formed by deleting the corresponding rows and columns associated with the zero entries.
In \eqref{eq:13}, the objective function is convex and all constraints are affine in $\mathbf{\tilde{Q}}$, 
so \eqref{eq:13} defines a convex optimization problem that can be efficiently solved using CVX \cite{cvx1}. 
The solution yields the matrix $\mathbf{\tilde{Q}}^\star \in \mathbb{C}^{N \times N}$. 
If $\operatorname{rank}(\mathbf{\tilde{Q}}^\star)=1$, the unconstrained DMA weights can be directly recovered from the dominant eigenvector of $\mathbf{\tilde{Q}}^\star$, since $\mathbf{\tilde{Q}}^\star = \mathbf{\tilde{q}}^\star \mathbf{\tilde{q}}^{\star H}$. 
Otherwise, a best rank-one approximation based on eigenvalue decomposition can be employed to obtain the DMA weights, as commonly adopted in the literature~\cite{aaltinokluLCH}%
\footnote{Although the SDP relaxation does not theoretically guarantee a rank-one solution, all simulations in this work resulted in rank-one optimal solutions.}.

Once the solution for the unrestricted DMA weights 
$\mathbf{\tilde{q}}^{\star}$ is obtained by solving the SDP problem in \eqref{eq:13}, 
it must be projected onto the Lorentzian circle defined in \eqref{eq:5}. 
This projection can be expressed through the mapping operator
\begin{equation}
\label{eq:gmlch}
\mathbf{q}^{\mathrm{LCH}} = \mathcal{M}\!\left(\mathbf{\tilde{q}}^{\star}; \Omega \right),
\end{equation}
where $\mathbf{q}^{\mathrm{LCH}} \in \mathbb{Q}^{N \times 1}$ and $\Omega$ denotes the chosen Lorentzian-constrained mapping scheme. 
For this operator, the recently proposed mapping method, namely the
Adaptive-Radius Lorentzian-Constrained Holography (ARLCH), is integrated
into the proposed beamforming algorithm for optimizing DMA weights
\cite{aaltinokluLCH}. In ARLCH, the diameter of the Lorentzian circle is
relaxed while determining the optimal phase points of the
Lorentzian-constrained DMA weights from the unconstrained solution. The
projection is performed at an adaptively optimal diameter (Lemma 1, \cite{aaltinokluLCH}) that minimizes
the discrepancy between the ideal unconstrained weights and the feasible
Lorentzian-constrained weights. Then, these optimal phase points are used to determine Lorentzian-constrained weights using unitary Lorentzian-circle \eqref{eq:5}. Overall, ARLCH\footnote{For further details and derivations, the reader is referred to \cite{aaltinokluLCH}.} enables dynamic phase
optimization on a relaxed Lorentzian circle with an adaptive radius, thereby providing additional degrees of freedom to control the sinusoidal variations imposed by Lorentzian constraints. In contrast, traditional Lorentzian projection schemes map the solution onto a unitary Lorentzian circle, which limits the flexibility of controlling amplitude variations of the DMA weights.
As baseline methods to ARLCH, we consider the following classical LCH schemes as a mapping operator $\mathcal{M}(\cdot;\Omega)$: LCPH, LCUSH, and AOH, described below.
Given the SDP solution of \eqref{eq:13} for tier-1 DMA weight optimization,
\(
q_n^\star = |q_n^\star| e^{j\psi_n^\star}, 
\quad \forall n \in \{1,\dots,N\},
\)
the Lorentzian-constrained DMA weights for conventional schemes are obtained as follows:
\begin{itemize}
    \item \textbf{LCPH}: 
    \(
    q_n^{\mathrm{LCH}} =
    \begin{cases}
    \sin(\psi_n^\star)e^{j\psi_n^\star}, & 0 \leq \psi_n^\star \leq \pi, \\[2pt]
    0, & \pi < \psi_n^\star \leq 2\pi,
    \end{cases}
    \)
    which preserves the SDP phase while imposing Lorentzian amplitude shaping.
    
    \item \textbf{LCUSH}:  
    \(
    q_n^{\mathrm{LCH}} = \tfrac{1}{2}\bigl(j+e^{j\psi_n^\star}\bigr),
    \)
    which maps the SDP phase onto a fixed Lorentzian circle.
    
    \item \textbf{AOH}:  
    \(
    q_n^{\mathrm{LCH}} = \sin(\theta_n)\, e^{j\theta_n},
    \)
    where \(\theta_n = \arcsin\!\left(\tfrac{1+\cos(\psi_n^\star)}{2}\right)\),
    which performs amplitude-optimized holographic mapping.
\end{itemize}

\begin{algorithm}[t]
\caption{Proposed Algorithm for Solving Problem \eqref{eq:10}}
\label{alg:AO_joint}
\begin{algorithmic}[1]
\STATE Initialize DMA weights $\mathbf{Q}^{(0)}$ randomly
\STATE \textbf{[Start]} Solve SDP \eqref{eq:12} for $\{\mathbf{w}_k^{(0)}\}$, $\{\rho_k^{(0)}\}$ given $\mathbf{Q}^{(0)}$
\FOR{$t = 1$ to $T_{\max}$}
   \STATE \textbf{Step 1: DMA Weight Update:}  
   Solve SDP \eqref{eq:13} for unconstrained $\tilde{\mathbf{q}}^{\star (t)}$
   \STATE Lorentzian projection:  
   $\mathbf{q}^{(t)} = \mathcal{M}\!\bigl(\tilde{\mathbf{q}}^{\star (t)};\Omega\bigr)$
   \STATE \textbf{Step 2: Precoder and Splitting Update:}  
   Solve SDP \eqref{eq:12} for $\{\mathbf{w}_k^{(t)}\}$ and $\{\rho_k^{(t)}\}$ given updated $\mathbf{Q}^{(t)}$
   \STATE Update transmit power $P_{\text{Tx}}^{(t)}$
   \STATE \textbf{Convergence Check:}  
   If $P_{\text{Tx}}^{(t)} \leq P_{\text{Tx}}^{(t-1)} $, update $P_{\text{Tx}}^{(f)}$,  $\{\mathbf{w}_k^{(f)}\}$, $\{\rho_k^{(*)}\}$, $\mathbf{Q}^\star$
\ENDFOR
\STATE \textbf{Output:} Optimized $\mathbf{Q}^{(f)}$, $\{\mathbf{w}_k^{(f)}\}$, and $\{\rho_k^{(f)}\} \forall k \in K$
\end{algorithmic}
\end{algorithm}
Following the above derivations, the solution procedure for Problem~\eqref{eq:10} is summarized in Algorithm~\ref{alg:AO_joint}.

\subsection{Discussion}

Algorithm~\ref{alg:AO_joint} alternates between two convex SDP subproblems: DMA weight optimization (Step~1) and joint precoder and power-splitting optimization (Step~2), both solvable optimally via interior-point methods. Although theoretical proof for rank-one optimality is not shown here for these steps, it should be noted that all numerical experiments yielded rank-one solutions. The convergence behavior is therefore primarily governed by the Lorentzian projection applied after the SDP relaxation in Step~1. This projection is intrinsic to DMA hardware due to Lorentzian amplitude–phase coupling and is also required in alternative formulations, such as manifold-based approaches (via LCUSH) \cite{DMA_BF1}. The proposed ARLCH scheme performs this projection via a bounded minimization with an adaptive Lorentzian radius, enabling flexible realization of different LCH schemes within a unified framework. While the projection step may temporarily violate SINR and EH constraints, feasibility is fully restored in Step~2 through re-optimization of digital precoders and power-splitting ratios. Moreover, the explicit convergence check in Algorithm~\ref{alg:AO_joint} guarantees monotonic non-increasing transmit power across iterations, ensuring convergence to a stable solution satisfying both SINR and EH constraints. The robustness of this two-tier SDP-based Lorentzian projection framework was previously validated in \cite{aaltinokluLCH} through comparison with a lower-complexity ADMM-based approach for the LCUSH realization, where a negligible optimality gap was observed under identical LCH settings. This demonstrates consistent convergence and solution accuracy, while enabling a broader class of LCH schemes. In the proposed algorithm, the computational complexity is dominated by SDP solvers in CVX, with order $\mathcal{O}\!\big(\max\{K,N\}^{4} N^{1/2}\log(1/\epsilon)\big)$, where $\epsilon$ denotes solver accuracy~\cite{5447068}. Although higher than ADMM or manifold-based methods, this framework enables fair comparison of multiple LCH schemes, including ARLCH, under identical PS-SWIPT settings and practical constraints such as nonlinear energy harvesting.
This SDP-based approach remains computationally feasible for moderate system sizes $(K,N)$ \cite{aaltinokluLCH}. Developing scalable, lower-complexity implementations remains an important direction for future work.

\section{NUMERICAL RESULTS}
\begin{figure}[!t]
\centering
\includegraphics[width=0.35\textwidth]{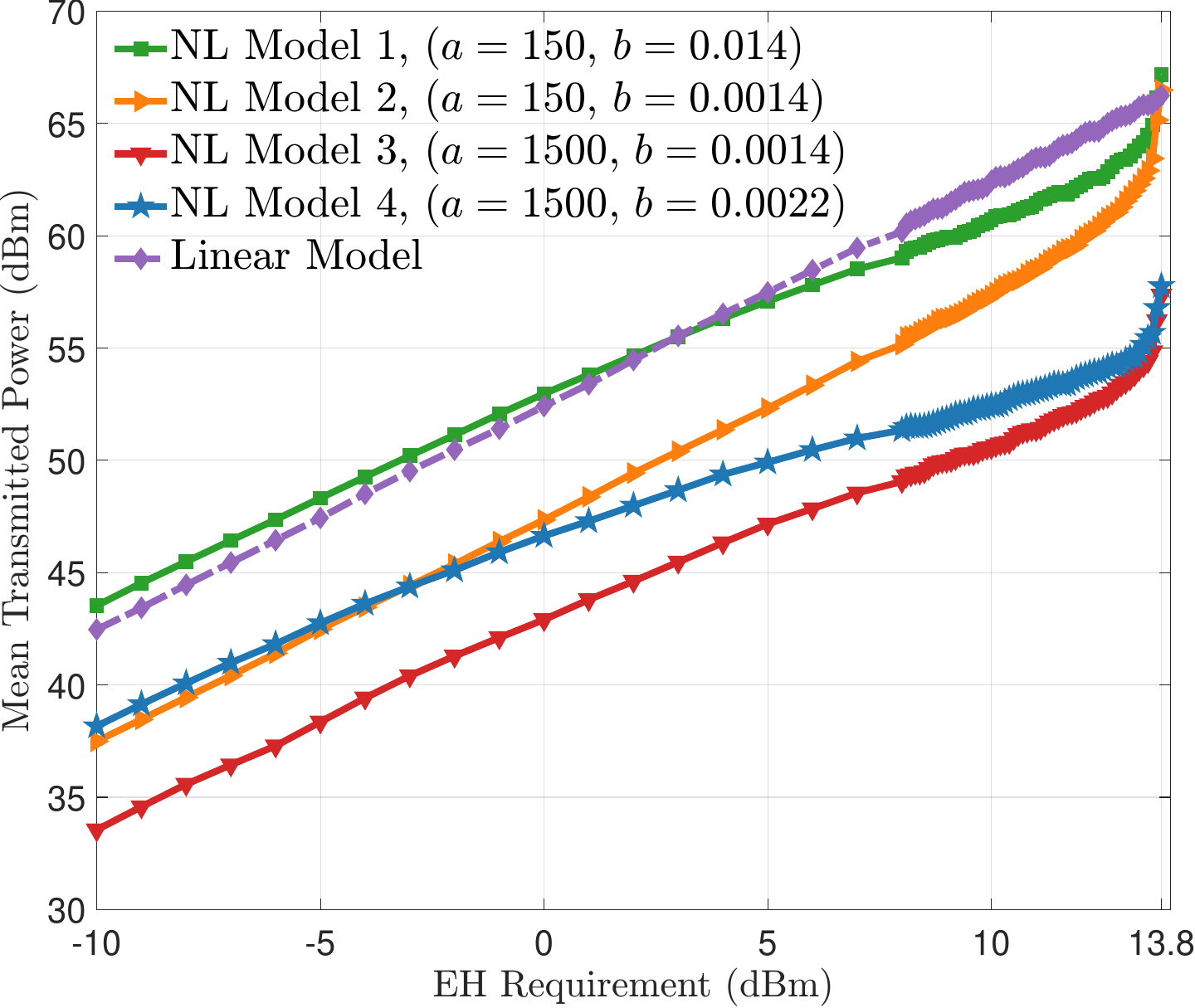}
\caption{Transmitted power versus EH power requirement \((E_k^{\mathrm{th}})\) for different energy harvesting models with \(K = 2\) and \(\delta_k=10~\mathrm{dB}\).}

\label{fig:results_1}
\end{figure}

In the numerical results, the DMA array is placed on the $xy$-plane, 
while the UEs are located in the near-field region on the $xz$-plane, 
with distances satisfying $0.1d_{\mathrm{F}}<r<d_{\mathrm{F}}$. Here, \(d_\mathrm{F}\) denotes the Fraunhofer distance, given by \(d_\mathrm{F} = \frac{2D^2}{\lambda}\)
where \(D\) and \(\lambda\) represent the antenna aperture length and wavelength, respectively. The carrier frequency is set to \(f = 28~\mathrm{GHz}\). The DMA parameters are selected for attenuation constant \(\alpha = 0.6~\mathrm{m}^{-1}\) and phase constant \(\beta = 827.67~\mathrm{rad/m}\), with inter-element spacing of \(\lambda/2\) along both \(x\)- and \(y\)-directions. The DMA configuration consists of \(N_\mathrm{r} = 8\) RF chains and \(N_\mathrm{e} = 64\) metasurface elements per RF chain, resulting in an effective aperture of \(D \approx 0.33~\mathrm{m}\) and a corresponding Fraunhofer distance of \(d_\mathrm{F} \approx 21.5~\mathrm{m}\). The antenna noise power is set to \(\sigma_a^2 = -70~\mathrm{dBm}\) in all simulations. 
As an initial analysis, we evaluate the performance of the proposed approach, combining ARLCH-based DMA mapping with optimal PS, under different NL energy harvesting models defined in~\eqref{eq:nonlinear_EH_models}, along with a linear EH reference. Four NL models are considered with parameters adopted from the literature~\cite{nonlinear_harvester}, all exhibiting a saturation power of \(E^{\text{sat}} = 13.8~\mathrm{dBm}\). The linear model assumes a constant conversion efficiency of \(\eta = 0.5\). The analysis is performed for a two-user scenario (\(K = 2\)), where both users are aligned in the same direction with UE$_1$ located at \((0, 0, 0.1d_\mathrm{F})\) and UE$_2$ at \((0, 0, 0.3d_\mathrm{F})\). In this setup, beamforming is achieved through near-field beam focusing based on the spherical-wave model. The results in Fig.~\ref{fig:results_1} depict the transmitted power versus the EH power requirement \(E_k^{\mathrm{th}}\) for a guaranteed SINR of \(\delta_k = 10~\mathrm{dB}\). As expected, the linear model shows a monotonic increase in transmitted power with \(E_k^{\mathrm{th}}\), whereas the NL models exhibit saturation behavior near \(E^{\text{sat}}\). Among the NL models, higher charging rate \(a\) values lead to lower transmitted power, while a smaller sensitivity parameter \(b\) further reduces the required transmit power for the same \(a\). The results demonstrate the robustness of the proposed beamforming approach in accurately capturing the nonlinear EH behavior with changing parameters, highlighting the importance of precise nonlinear modeling for reliable performance prediction.
\begin{figure}[!t]
\centering
\includegraphics[width=0.35\textwidth]{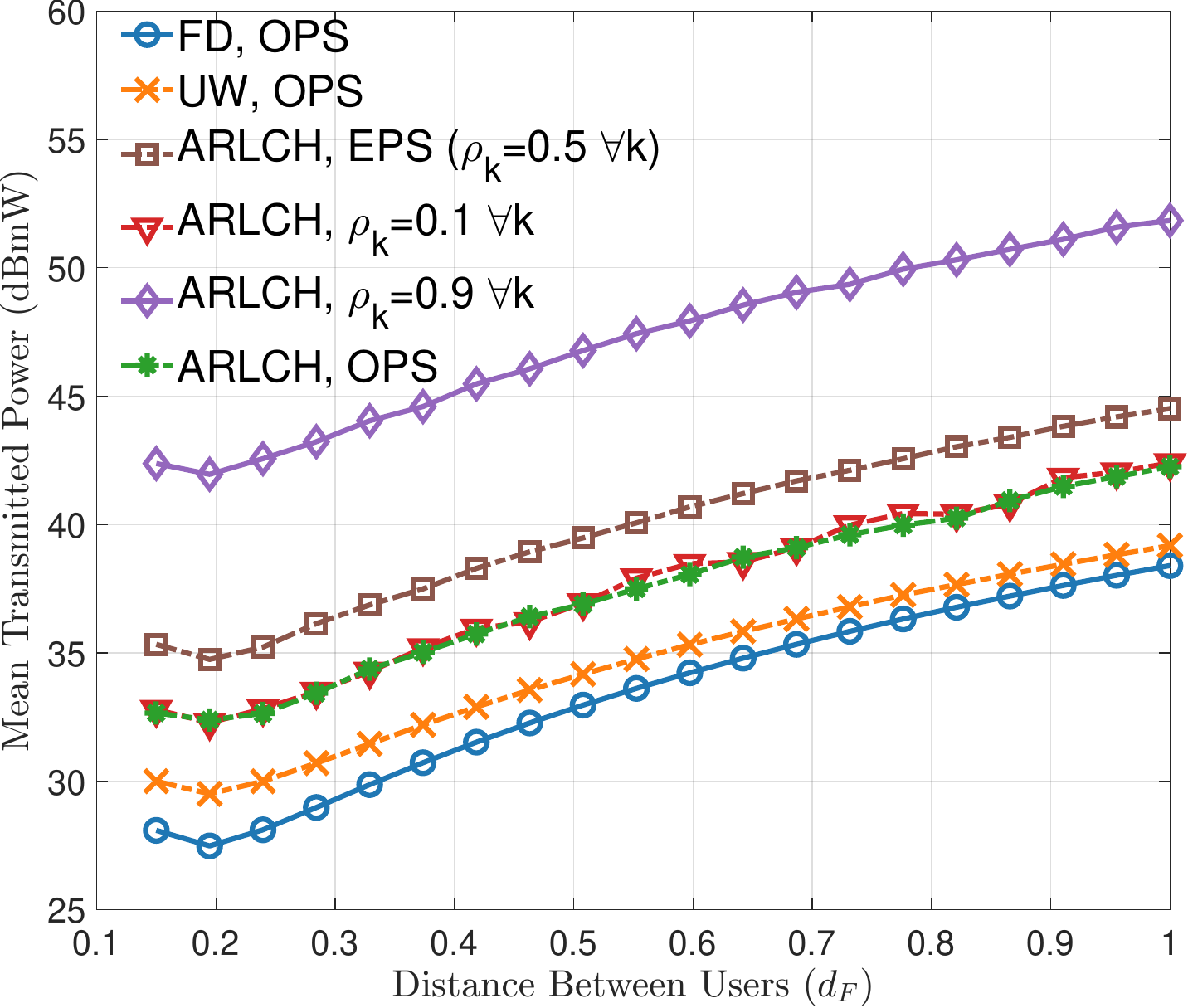}
\caption{Transmitted power vs. user separation distance for \(K=2\), \(\delta_k=10~\mathrm{dB}\), \(E_k^{\mathrm{th}}=-10~\mathrm{dBm}\), and \(\sigma_{\mathrm{c}}=-50~\mathrm{dB}\).}
\label{fig:results_2}
\end{figure}
\begin{figure}[!t]
\centering
\includegraphics[width=0.35\textwidth]{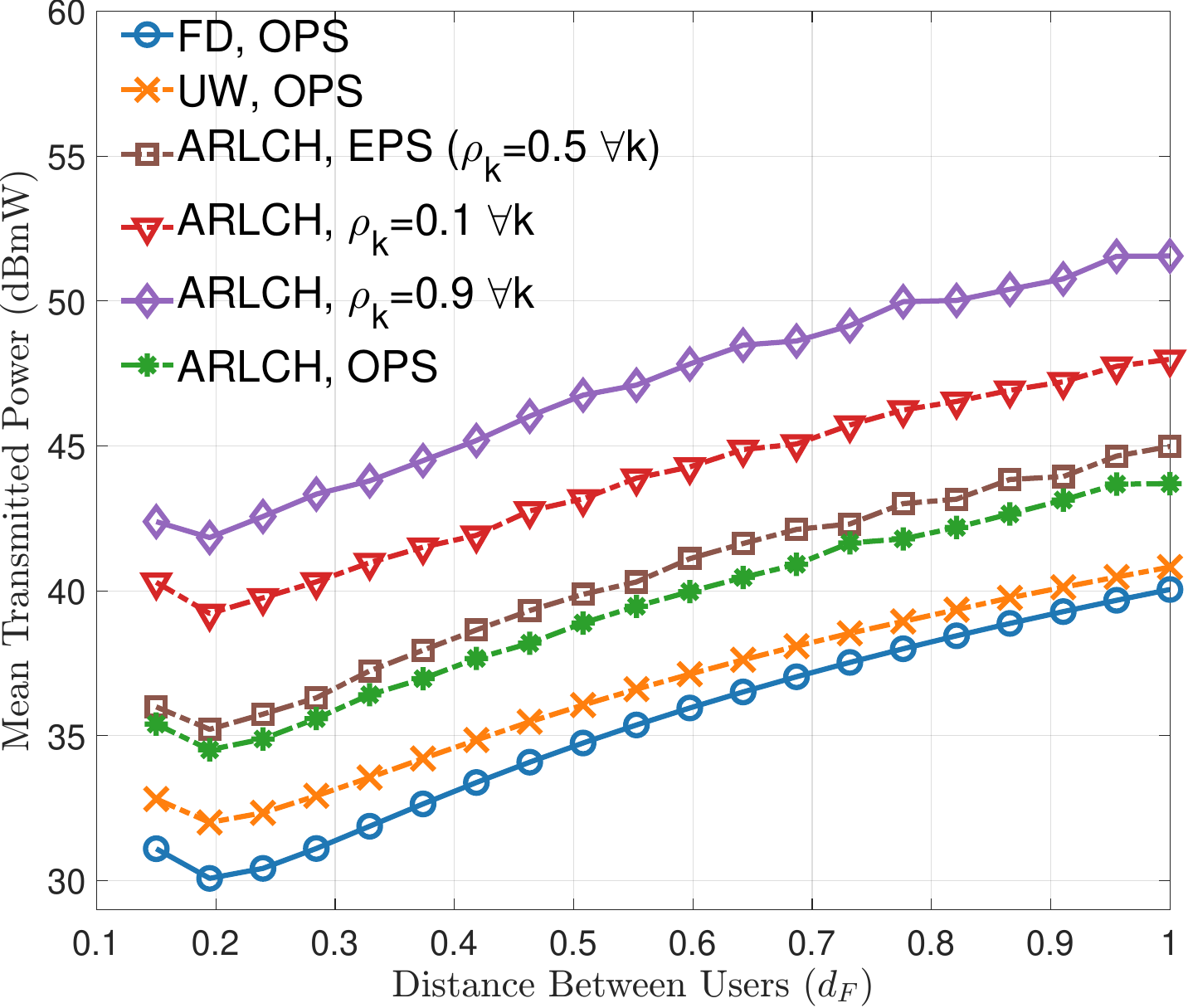}
\caption{Transmitted power vs. user separation distance for \(K=2\), \(\delta_k=10~\mathrm{dB}\), \(E_k^{\mathrm{th}}=-10~\mathrm{dBm}\), and \(\sigma_{\mathrm{c}}=-30~\mathrm{dBm}\).}

\label{fig:results_3}
\end{figure}

We next examine the effect of PS and baseband conversion noise on the proposed ARLCH-based DMA beamforming with optimal PS (OPS). A two-user scenario (\(K=2\)) is considered, where UE$_1$ is fixed at \((0,0,0.1d_\mathrm{F})\) and UE$_2$ is swept along the $z$-axis as \((0,0,\zeta d_\mathrm{F})\), with \(\delta_k=10~\mathrm{dB}\) and \(E_k^{\mathrm{th}}=-10~\mathrm{dBm}\). As baselines, we include the FD case (\(64\times8\) array with 512 RF chains) and the DMA-UW. Equal power splitting (EPS) and fixed splitting (\(\rho_k=0.1,0.9\)) are compared with OPS under \(\sigma_{\mathrm{c}}=-50\) and \(-30~\mathrm{dBm}\) in Fig.~\ref{fig:results_2}, and Fig.~\ref{fig:results_3}, respectively.
The results show that as UE$_2$ moves farther from the DMA, the transmitted power increases in all cases, since satisfying the EH requirement becomes more challenging at larger distances. When UE$_2$ moves closer than \((0, 0, 0.2d_\mathrm{F})\), increased inter-user interference raises the required transmit power, despite ease of EH requirement guarantee with a lower path-loss. At lower conversion noise (\(\sigma_{\mathrm{c}} = -50~\mathrm{dB}\)), where the SINR constraint is easier to meet, the OPS performance converges toward \(\rho_k = 0.1\) performance, indicating WPT dominance, as shown in Fig.~\ref{fig:results_2}. Conversely, at high noise (\(-30~\mathrm{dB}\)) it shifts toward EPS with balanced WIT–WPT tradeoff as depicted in Fig.~\ref{fig:results_3}. These results confirm the effectiveness of the proposed OPS-based beamforming for adaptive power allocation under varying channel and hardware conditions. Finally, comparison with DMA-UW and FD architectures illustrates the impact of the Lorentzian constraint: while amplitude–phase coupling in LCH significantly limits performance, the small performance gap between DMA-UW and FD validates the robustness of the proposed SDP-based optimization framework.

\begin{figure}[!t]
\centering
\includegraphics[width=0.35\textwidth]{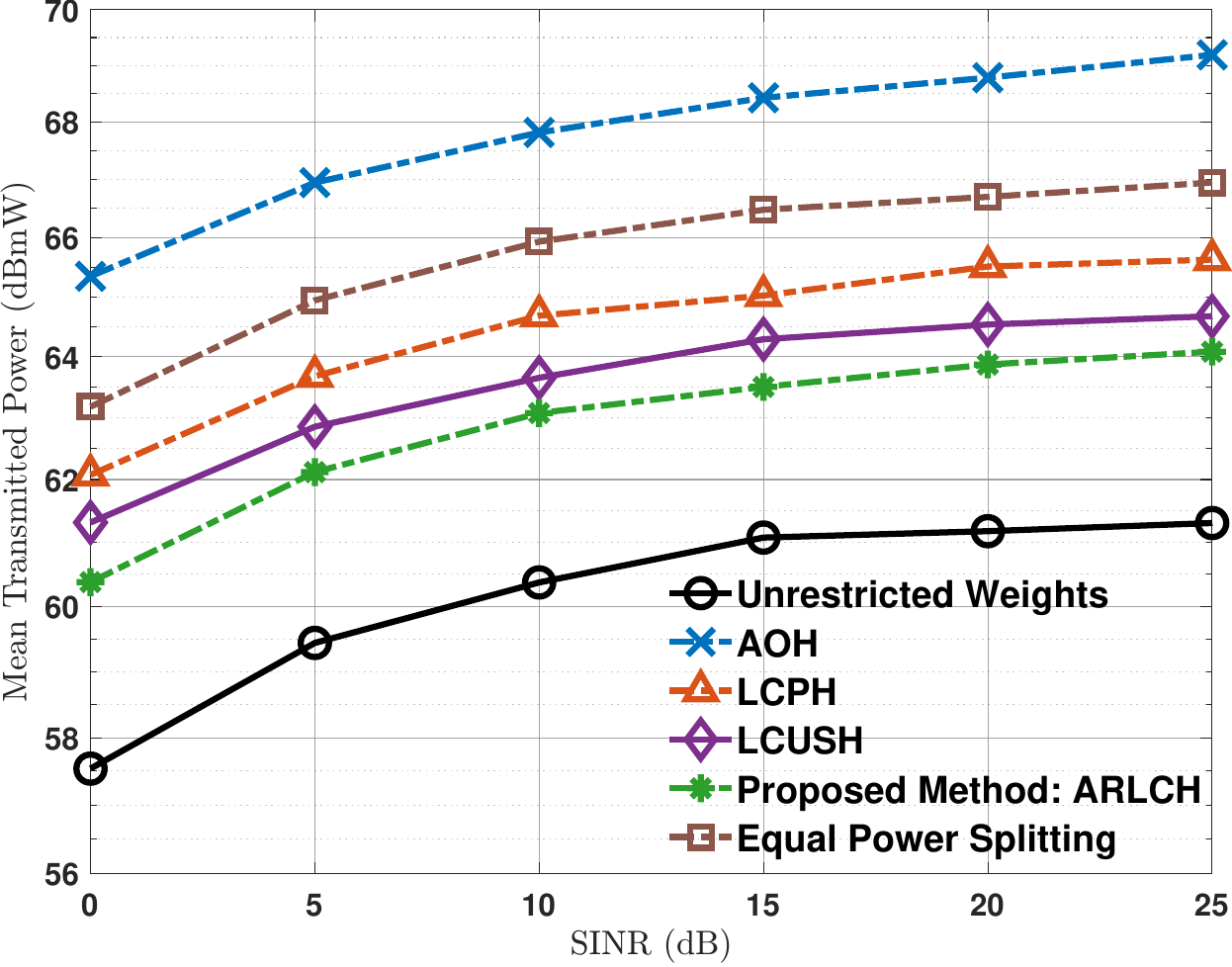}
\caption{Mean transmitted power vs. SINR for Monte Carlo realizations under different LCH schemes with \(E_k^{\mathrm{th}} = -10~\mathrm{dBm}\) and \(K = 4\).}

\label{fig:results_4}
\end{figure}

To examine further the impact of different LCH schemes, Monte Carlo simulations with 200 realizations are conducted for AOH, LCPH, LCUSH, and DMA-UW, alongside the proposed ARLCH approach and its EPS variant. Fig.~\ref{fig:results_4} illustrates the mean transmitted power versus SINR threshold~$\delta_k$. Results indicate that AOH performs poorly due to its amplitude-only optimization, failing to achieve proper phase alignment, while LCPH also degrades performance since significant number of the DMA elements are deactivated to preserve phase matching. LCUSH improves performance slightly by activating more elements, yet the proposed ARLCH achieves the best performance by adaptively balancing amplitude-phase relations, reducing grating lobes, and lowering transmit power while maintaining SINR targets. Conversely, EPS under ARLCH significantly degrades performance, emphasizing the importance of optimal power splitting in PS-SWIPT systems. These results clearly demonstrate the robustness of proposed approach and also highlight the significance of selection of proper LCH scheme.
\section{CONCLUSION}

This paper presents an SDP-based framework for optimal PS-SWIPT in DMA-assisted multiuser MISO systems. A joint optimization of DMA weights, precoding, and PS ratios is achieved through an alternating optimization algorithm with Lorentzian-mapping scheme based on ARLCH. The results confirm the critical role of mapping schemes, which are unique to DMA architectures, in conjunction with optimal PS for effectively balancing the EH–ID trade-off. The results demonstrate that the proposed approach effectively captures nonlinear EH characteristics, minimizes transmit power, and outperforms conventional LCH. Future work can explore its extension to the non-line-of sight scenario with electromagnetic compliant modeling of DMA.

\bibliographystyle{IEEEtran}
\bibliography{journal_abbreviations,references}

\end{document}